\documentclass[11pt] {article}      
\usepackage{amsmath,amssymb, amsthm, eucal,accents}  
\usepackage{latexsym, graphicx, amscd}
\usepackage{newtxtext,newtxmath}      
\usepackage[T1]{fontenc}     
\usepackage{enumerate}       
\usepackage{float}                 
\usepackage{mathtools}
\usepackage{tikz}
\usetikzlibrary{matrix, arrows, calc, fit}
\usetikzlibrary{arrows, decorations.markings}
\tikzstyle{AArrow} = [thick, decoration={markings,mark=at position 1 with {\arrow[semithick]{open triangle 60}}},%
   double distance=1.4pt, shorten >= 5.5pt,
   preaction = {decorate},
   postaction = {draw,line width=1.4pt, white,shorten >= 4.5pt}]
\tikzstyle{AArroww} = [semithick, white,line width=1.4pt, shorten >= 4.5pt]
\usepackage[strict]{changepage}
\usepackage{relsize}
\theoremstyle{plain}
\newtheorem{theorem}{Theorem}[section]            
\newtheorem{proposition}[theorem]{Proposition}  

\theoremstyle{definition}
\newtheorem{definition}[theorem]{Definition}

\numberwithin{theorem}{section}
\numberwithin{equation}{section}
\newcommand{\gaction}[2]{\genfrac{}{}{0.5pt}{}{#1}{#2}%
                        \!\lower2pt\hbox{\rotatebox[origin=c]{-90}{{$\looparrowright$}}}}

\newcommand{\dotfraction}[2]{\genfrac{}{}{0.5pt}{}{#1}{#2}%
                        \!\lower.5pt\hbox{{$\circ$}}}
\def\rmspan{\hbox{span}\,}

\def\rmspan{\hbox {\rm span}}

\newcommand{\ooverline}[1]{\overline{\overline{#1}}}
\renewcommand\appendix{\par
  \setcounter{section}{0}
  \setcounter{subsection}{0}
  \setcounter{figure}{0}
  \setcounter{table}{0}
  \renewcommand\thesection{Appendix \Alph{section}}
  \renewcommand\thefigure{\Alph{section}\arabic{figure}}
  \renewcommand\thetable{\Alph{section}\arabic{table}}
}
\usepackage{titlesec}                                                           
\titlelabel{\thetitle.\ }                                                           
\titleformat*{\section}{\fontsize{14pt}{14pt} \bf}                   
\usepackage{fancyhdr}
\usepackage{sidecap}  
\usepackage[hang,small,sc]{caption}

\def\QED{ $\square$}

\def\ee{\varepsilon}
\def\ff{\varphi}

\def\moplus{\boxplus}
\def\voplus{\oplus}

\def\rmid{\hbox{\rm id}}
\def\rmid{\hbox{\rm id}}

\def\hhat#1{\widehat{#1}}

\begin{document}

\title{\bf Making sense of~relativistic composition \\  of velocities}
\author{Jerzy Kocik                  
\\ \small Department of Mathematics
\\ \small Southern Illinois University, Carbondale, IL62901
\\ \small jkocik{@}siu.edu  
}
\date{}

\maketitle

\begin{abstract}
\noindent
What does it mean to ``add'' velocities relativistically --
clarification of the conceptual problems,
new derivations of the related formulas, 
and identification of the  source of the non-associativity of 
the standard vector version of the addition formula
are addressed.
\\[3pt]
\\[3pt]
\end{abstract}


\section{ Introduction}

Composition of velocities, also nicknamed as ``addition of velocities'', 
is somewhat confusing due to unclear concept of velocity 
as read in different frames of reference in the contexts of space-time.
Recall that in the case of collinear velocities, Poincare formule \cite{Po} formula of 1904 holds:
\begin{equation}
\label{eq:1}
\mathbf v\oplus \mathbf u = \frac{\mathbf u +\mathbf v}{1+\mathbf u \mathbf v}
\end{equation}
It is a well-behaved property, defining a group on the interval $-1,\,1]$.
However, in the case the velocities are not collinear, the rule assumes a rather horrifying form
\begin{equation}
\label{eq:2}
\mathbf v\oplus \mathbf u   
= \frac{    \sqrt{1-|\mathbf v|^2} \, \mathbf u  +
                    \left( \frac{ \left( 1-\sqrt{1-|\mathbf v|^2} \right)           \, \mathbf v\cdot \mathbf u }
                                {|\mathbf v|^2}
                                                        +1 \right)\, \mathbf v
}
      {1 + \mathbf v\cdot \mathbf u}
\end{equation}
Were did the elegance \eqref{eq:1} go?
Not only does  the algebraic form looks unattractive 
but also the ``product'' is noncommutative and even nonassociative,
very much at odds with our common Galilean intuition. 
This is in contrast with associativity of the  Lorentz group.
As an escape, most (if not all) textbooks do not report \eqref{eq:2}, 
not to mention deriving it,
and jump immediately to the group description.
\\

In these notes  we derive the formula with geometry of space-time (no reference to Lorentz group),
clarify the meaning of thre ``addition of velocities'',  
and explicate the origin of the confusion over nonassociativity.

\section{Observers in the Minkowski space}

Let $M$ be the Minkowski space, a vector space equipped with an inner product $\langle \cdot,\cdot\rangle$ of signature $(1,n)$,
one ``plus'' and $n$ ``minuses''. 
(The restriction to $n=3$, standard in basic physics, is not essential here.) 
The product defines orthogonality of vectors: $\mathbf A\bot \mathbf B$ iff $\langle \mathbf A,\mathbf B\rangle =0$.  
A future-oriented unit vector $\mathbf T\in M$ is called vector {\bf chronor}.
It determines a space-like subspace
\begin{equation}
\label{eq:space}
\mathbf T^\bot = \{\,\mathbf x \in M \;|\; \mathbf x \bot \mathbf T\,\}
\end{equation}
Every split of $M$ into the direct sum  
$$
M= \mathbf T\ \oplus \ \mathbf T^\bot \
$$
is tantamount to an {\bf observer}.
The subspace $\mathbf T^\bot$ is called the {\bf private space} of the observer $\mathbf T$
(short for instantaneous observer's space).
The set of all chronors will be denoted by $\mathcal T$.
It is topologically equivalent to the upper piece of the unit hyperboloid,  
 $\mathcal T  \cong \mathbb R^n$. 
\\
\\
{\bf Notation:}  For  $\mathbf A\in M$:  $\|\mathbf A\|^2 = \langle\mathbf A,\,\mathbf A\rangle$.  
For space-like vectors we use alternative  notation:
$$
  \mathbf a\cdot \mathbf b = - \langle\mathbf a,\,\mathbf b\rangle
\qquad
  |\mathbf a|^2 = - \|\mathbf a\|^2 \ \geq 0
$$

The Lorenz group $\Lambda$ is the connected component of the group of isometries of $M$. 
It usually dominates presentations of relativity as a convenient tool to get results quickly. 
Our goal is, however,  to stay in the framework of pure geometry as long as possible.
\\
\\

The ``addition of two velocities'' is understood as a way to determine the 
mutual relation of two observers (chronors), given mutual relations between each of them and another observer.
In Galilean physics, this was modeled by the usual sum of vectors in space.
In relativity, due to the extension of the space to space-time, things became less obvious.
In the following, 
we provide a number of ways how the concept of ``addition of velocities'' may be adjusted to this new environment.
\begin{figure}[h]
$$
\begin{tikzpicture}[baseline=-0.7ex]
    \matrix (m) [ matrix of math nodes,
                         row sep=3.7em,
                         column sep=4em,
                         text height=3ex, text depth=2ex] 
 {
  ~ &\quad \mathbf T_2 \quad    &   \quad   \\
   \mathbf T\equiv \mathbf T_1 &   ~  & \quad \mathbf T_3 \quad    \\
  };
    \path[-stealth]
       (m-2-1) edge [transform canvas={yshift=0.3ex}] node[above] {$\mathbf v_{13}$} 
                                                                              node[below]  
 {$  \mathbf v\hhat\oplus \mathbf w    \ = \ \mathbf v \oplus \mathbf u  $}     
(m-2-3)
        (m-2-1) edge 
                            node[left] {$\mathbf v $~~~} 
                            node[right] {~~$\mathbf v_{12}~~~$}              (m-1-2)
        (m-1-2) edge node[right]   {~~$\mathbf w$}
                            node[left] {$\mathbf v_{23}$~~}        (m-2-3)
;
\end{tikzpicture}   
\quad 
\mathbf u = \hbox{adjusted } \mathbf w
$$
\caption{Notations for the velocities between three objects / observers}
\label{fig:vvv}
\end{figure}
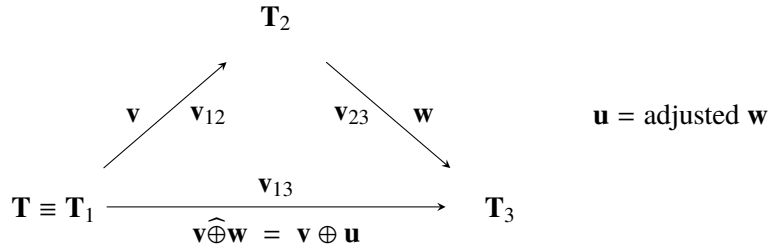

\section{Groupoid of chronons (observers) }

\begin{figure}[h]
\centering
\includegraphics[scale=.89]{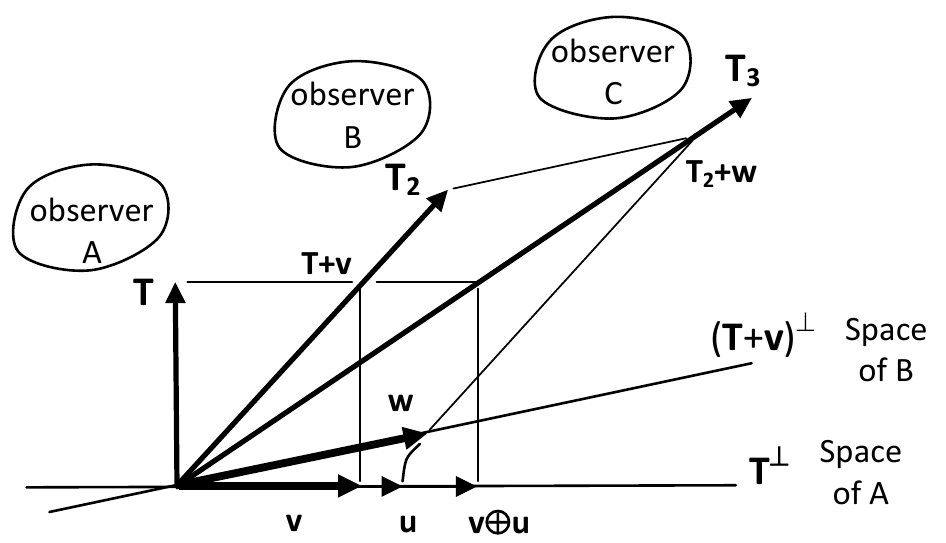}
\caption{\small Three observers) and their relations.
Not to scale. Also, $\mathbf v$ and $\mathbf u$ do not need to be collinear}
\label{fig:TTT}
\end{figure}

The most elementary expression of the mutual relation between two observers represented by chronors $\mathbf T_1$ and $\mathbf T_2$
is simply pair
$(\mathbf T_1, \mathbf T_2)$.
All such pairs form a {\bf groupoid} the underlying set of which is topologically equivalent to 
$\mathcal T\times\mathcal T$ and the product is defined by 
$$
(\mathbf A, \mathbf B)\circ (\mathbf C, \mathbf  D) = (\mathbf A, \mathbf D)
\qquad
\hbox{if}\quad
\mathbf B = \mathbf C
$$
(if $\mathbf B\not= \mathbf C$, the product is not defined).
Thus we may say that the following relation is the most abstract 
version of the formula we are seeking:
\begin{equation}
\label{eq:add1}
\boxed{
\begin{array}{cc} \hbox{\small \sf Velocity addition}\\[-3pt] \hbox{\small \sf formula nr 1} \end{array} \qquad
(\mathbf T_1, \mathbf T_2)\circ (\mathbf T_2, \mathbf  T_3) = (\mathbf T_1, \mathbf T_3)
\qquad~~
}
\end{equation}
The matter is in how we measure the mutual configuration between two observers, i.e. chronors.
This brings us to the notion of {\bf velocity}.
We say that an object with chronor $\mathbf T_2$ has velocity $\mathbf v_{12}\in \mathbf T_1^\bot$ 
with respect to observer $\mathbf T_1$ if
$$
\mathbf T_2\, \|\, (\mathbf T_1+\mathbf v_{12} )
$$
(Symbol $\|$ denotes parallel. See Figure \ref{fig:TTT}.) 
It must be stressed that so-called reciprocal velocities are not mutual negatives.  
They belong to different space-like subspaces and in general $\mathbf v_{12}+\mathbf v_{21}\not= \mathbf 0$. 

It will be convenient to have a symbol for normalizing a vector.
Define 
$$
\ooverline{\mathbf A} \  = \ \frac{\mathbf A}{|\mathbf A|} \,.
$$
Now, we may express the definition of velocity as  
\begin{equation}
\label{eq:def2}
\mathbf T_2 = \ooverline{\mathbf  T_1+\mathbf v_{12}}
\end{equation}
Equation \eqref{eq:add1} reads now
$$
\label{eq:}
(\mathbf T_1 ,\; \ooverline{\mathbf T_1+\mathbf v_{12}}  )
\ \circ \
(\ooverline{\mathbf T_1+\mathbf v_{12}}, \;     \ooverline{\ooverline{\mathbf T_1+\mathbf v_{12}}   + \mathbf v_{23}} ) 
\ \ = \ \ 
(\mathbf T_1 ,\; \ooverline{\mathbf T_1+\mathbf v_{12}\!\oplus\!\mathbf v_{23}} ) 
$$
which may simply be reduced to
%
%
\begin{equation}
\label{eq:add2}
\boxed{
\begin{array}{cc} \hbox{\small \sf Velocity addition}\\[-3pt] \hbox{\small \sf formula nr 2} \end{array} \qquad
\ooverline{\ooverline{\mathbf T+\mathbf v}   + \mathbf w}  
\ \ = \ \ 
 \ooverline{\mathbf T+\mathbf v\hhat\oplus\mathbf w}  
\qquad~~}
\end{equation}
The symbol for the composition, $\hhat\oplus$ is decorated by a hat to indicate that our formula 
concerns the map
$$
\hhat{\oplus} : \ \mathbf T^\bot \times( \mathbf T+\mathbf v)^\bot \ \to \ \mathbf T^\bot
$$
defined for pairs of vectors from {\it different} space-like subspaces of $M$.
This will be fixed later.

Let us also introduce a aan observer-indepedent scalar measure of the mutual configuration of two chronors.
Denote $v= |\mathbf v_{12}| = |\mathbf v_{21}|$ and let us define ``slowness'':
$$
\delta = \quad |\mathbf T_1+\mathbf v_{12}| 
$$
Then the following can be easily verified:
\begin{equation}
\label{eq:delta}
\delta = \sqrt{1-v^2}
\qquad
\langle \mathbf T_1,\,\mathbf T_2\rangle = \frac{1}{\delta}
\qquad
\| \mathbf T_1 + \mathbf T_2 \|^2 = 2\frac{1+\delta}{\delta}
\end{equation}
In physics literature, a more popular is the reciprocal entity: $\gamma = 1/\delta$.

\section{Velocity addition via pure geometry}

\begin{figure}[b]
\centering
\includegraphics[scale=.97]{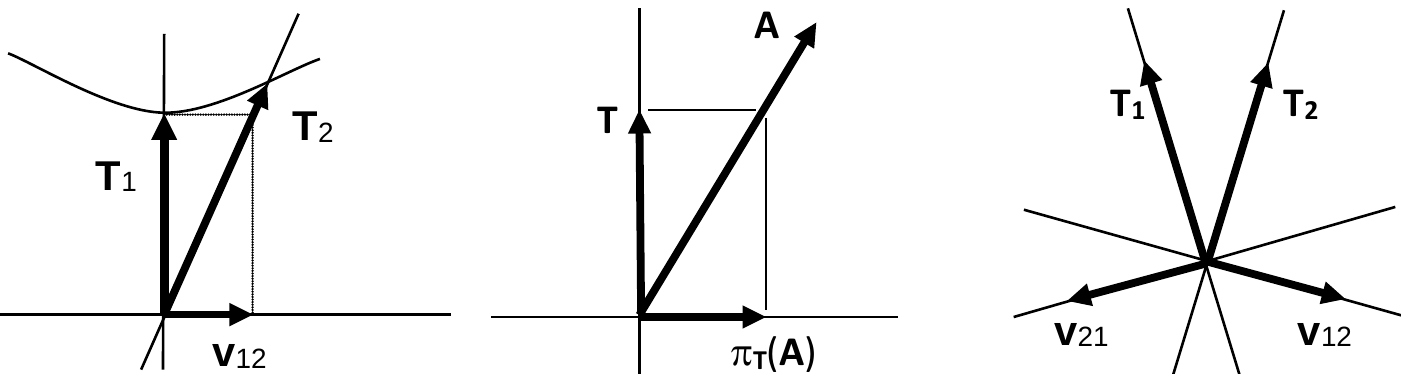}
\caption{\small {\bf Left:} Definition of velocity. {\bf Center:} Definition of    pseudo-projection.\\
                      {\bf Right:} Reciprocal velocities.} 
\label{fig:ppp}
\end{figure}

Let us start with a more direct formula for the velocity.
Suppose a vector $\mathbf A\in M$ is given.  
We define a {\bf pseudo-projection} along $\mathbf T\in \mathcal T$ as a map
\begin{equation}
\label{eq:projection1}
\pi_{\mathbf T}: M\longrightarrow \mathbf T^\bot
\end{equation}
such that 
\begin{equation}
\label{eq:projection2}
\left(\mathbf T +\pi_{\mathbf T}(\mathbf A)\right) \wedge \mathbf A = 0
\end{equation}
A quick analysis of Figure \ref{fig:ppp} gives the explicit expression for the pseudo-projection:
\begin{equation}
\label{eq:projection3}
\pi_{\mathbf T}(\mathbf A) = \frac{\mathbf A}{\langle \mathbf A, \mathbf T \rangle} - \mathbf T
\end{equation}
One can easily check that:
$$
\begin{array}{cll}
(i) &  \pi_{\mathbf T}(\mathbf A) \ \in \rmspan \{\mathbf A,\, \mathbf T\} 
     &\qquad\hbox{same  as \eqref{eq:projection2}}  \\
(ii) & \pi_{\mathbf T}(\mathbf A) \in \mathbf T^\bot  
      &\qquad\hbox{i.e.,}\ \langle  \pi_{\mathbf T}(\mathbf A),\, \mathbf T\rangle = 0  
\end{array}
$$

Note that the definition \eqref{eq:projection3} is independent of scaling of $\mathbf A$.
Hence $\pi_{\mathbf T}(\mathbf A)$ is determined by {\it direction} of $\mathbf A$ in $M$ solely.
We now have the following convenient definition:

\begin{definition}
The {\bf velocity} of observer described by chronor $\mathbf T'$ with respect to an observer $\mathbf T$ is defined as a vector
$$
\mathbf v = \pi_{\mathbf T}(\mathbf T') \quad \in \mathbf T^\bot
$$    
\end{definition}

Clearly, the elements of the groupoid \eqref{eq:add1} are pairs of type: 
$$
(\mathbf T, \ \ooverline{\mathbf T+ \mathbf v}\ )   \ \sim \  (\mathbf T, \ \mathbf T+ \mathbf v) 
$$
where the second entry does not need to be normalized.  
\\

\noindent
{\bf Adding velocities via geometry.}
Suppose we have three observers defined by (normalized) chronors
$\mathbf T \equiv \mathbf T_1$, $\mathbf T_2$ and $\mathbf T_3$.
(See Figure \ref{fig:TTT}.)
Denote velocities:
$$
\begin{array}{cll}
\mathbf v&=\pi_{\mathbf T_1}(\mathbf T_2)    &  \in \mathbf T_1^\bot  \\
\mathbf w &= \pi_{\mathbf T_2}(\mathbf T_3)   &   \in \mathbf T_2^\bot \\
\mathbf v \,\hhat\oplus\, \mathbf w &= \pi_{\mathbf T_1}(\mathbf T_3)  &    \in \mathbf T_1^\bot
\end{array}
$$
(consult Figure \ref{fig:vvv}).
The goal is to express $\mathbf v \hhat\oplus \mathbf w$ in terms of $\mathbf v$ and $\mathbf w$.
Here is the answer:

\begin{equation}
\label{eq:add3}
\boxed{
\begin{array}{cc} \hbox{\small \sf Velocity addition}\\[-3pt] \hbox{\small \sf formula nr 3} \end{array} \qquad\qquad
\mathbf v\hhat \oplus\mathbf w = \pi_{\mathbf T}(\ooverline{\mathbf T+\mathbf v}+\mathbf w)
\qquad~~
}
\end{equation}

~

\noindent
The explicit algebraic expression corresponding to the above follows.

\begin{proposition}  {\bf [Relativistic velocity addition -- week version]}  
With the above assumptions, the addition () resolves to the following algebraic equation:
\begin{equation}
\label{eq:add4}
\boxed{
\begin{array}{cc} \hbox{\small \sf Velocity addition}\\[-3pt] \hbox{\small \sf formula nr 4} \end{array} \qquad\qquad
\mathbf v \hhat \oplus \mathbf w
=\displaystyle\frac{\mathbf v  + \delta\,\mathbf w -\delta\,\langle\mathbf T, \mathbf w\ \rangle \mathbf T}
                            {1+ \delta \; \langle\mathbf T, \mathbf w\rangle}
\qquad~~
}
\end{equation}
where $\delta= \|\mathbf T +\mathbf v\| = \sqrt{1+|\mathbf v|^2} =\langle\mathbf T,\, \mathbf T_2\rangle^{-1}$.
\end{proposition}
\noindent
{\bf Proof:}
Measure the velocity for chronor $\mathbf T_3$ (skip unnecessary normalization) with respect to 
observer $\mathbf T_2$:
$$
\begin{array}{lll}
\mathbf v \hhat \oplus \mathbf w
              &=& \pi_{\mathbf T}(\mathbf T_3)  
              \ = \  \pi_{\mathbf T} \left( \displaystyle\frac{\mathbf T + \mathbf v}{|\mathbf T +\mathbf v|}+ \mathbf w \right)\\[14pt]
              &=& \displaystyle \frac{ \displaystyle\frac{\mathbf T + \mathbf v}{|\mathbf T +\mathbf v|}+\mathbf w}
                             { \displaystyle\frac{1}{|\mathbf T\! +\!\mathbf v|} + \langle\mathbf T, \mathbf w\rangle}-\mathbf T \\[29pt] 
              &=& \displaystyle \frac{\mathbf v  + |\mathbf T\!+\!\mathbf v|\,\mathbf w -|\mathbf T \!+\! \mathbf v|\,\langle\mathbf T, \mathbf w \rangle \,\mathbf T}
{1+ |\mathbf T\!+\!\mathbf v| \; \langle\mathbf T, \mathbf w\rangle}\\
\end{array}
$$
\QED

\newpage

\section{Adding velocities in a single private space}

Here is the problem: 
Although the addition \eqref{eq:add4} is well-defined and makes perfect sense, 
note that the vectors of velocities are in different space-like subspaces.
In order to have a product $\oplus$ well defined in a single space, namely a map:
$$
\oplus:\ \mathbf T^\bot\times\mathbf T^\bot \longrightarrow \mathbf T^\bot
$$ 
we need to map the second component $\mathbf w$ to the private space of the original observer, $\mathbf T^\bot$.
We will achieve it by a map that isometrically  turns $\mathbf T_1^\bot \equiv \mathbf T^\bot$ to $\mathbf T_2^\bot \equiv (\mathbf T+\mathbf v)$
so that $\mathbf T_1$ is mapped to $\mathbf T_2$ and any vector perpendicular to the plane $\rmspan\{\mathbf T_1, \mathbf T_2\}$ is preserved.
Denote such a map
$$
\xi_{12}:  \mathbf T_1^\bot \ \to \ \mathbf T_2^\bot \qquad\hbox{and}\qquad    \xi_{21}:  \mathbf T_2^\bot \ \to \ \mathbf T_1^\bot
$$
Clearly, $\xi_{12}\xi_{21} = \rmid$ \ (identity on the ``home space'', $\mathbf T^\bot$).


%

\begin{proposition} 
\label{thm:turn}
Let $\mathbf T_1\equiv \mathbf T$ and $\mathbf T_2 \sim \mathbf T^\bot$ be two chronors.  The image of a space-like vector $\mathbf a\in \mathbf T^\bot$ 
rotated to the space $\mathbf T_2^\bot$ is  
\begin{equation}
\label{eq:turn}
\mathbf a' = \xi_{12}(\mathbf a) =\mathbf a  - \frac{\langle\mathbf v, \mathbf a\rangle}
                            {\delta}                                 \mathbf T
- \frac{\langle\mathbf v, \mathbf a\rangle}
                            {\delta (\delta+1)}                                 \mathbf v
\end{equation}
\end{proposition}

\noindent
{\bf Proof:}
Assume $\mathbf a'$ is 
$$
\mathbf a' = \mathbf a + \alpha \mathbf T + \beta \mathbf a,
 \qquad \alpha,\beta \in \mathbb R
$$
Now, since $(\mathbf T+\mathbf v) \bot \mathbf a$, and $\|\mathbf a' \|^2=\|\mathbf a\|^2$, we get
$$
\begin{cases}
\alpha +\langle \mathbf v, \mathbf a\rangle + \beta \|\mathbf v\|^2 =0\\
\alpha^2 + \beta^2 \|\mathbf v\|^2 + 2\beta \langle \mathbf a, \mathbf v\rangle = 0
\end{cases}
$$
Solving for $\alpha$ and $\beta$ gives the claim.
\QED

~

Thus the idea is to represent $\mathbf v$ of any of the formulas presented so far as  an image of some $\mathbf u\in \mathbf T^\bot$
via  the map \eqref{eq:turn}
\begin{definition}
The {\bf M{\o}ller loop} is the loop  $(\mathbf T^\bot, \oplus)$  where the product 
$\oplus: \mathbf T^\bot\times\mathbf T^\bot \to \mathbf T^\bot$  is defined as
\begin{equation}
\label{eq:defoplus}
\mathbf v \oplus \mathbf u \  \ = \ \ \mathbf v \,\hhat\oplus \, \xi(\mathbf u)
\end{equation}
\end{definition}

\noindent
{\bf Remark:}
The pair $(\mathbf T^\bot,\, \oplus)$ has the neutral element, $0\oplus \mathbf v = 0\oplus \mathbf v = 0$, and it has inverse elements:
$\mathbf v\oplus (-\mathbf v)$.  It is however non-associative, hence not a group
(see \cite{Sb}). 
The name ``M{\o}ller loop'' was suggested by Zbigniew Oziewicz \cite{O}.

\newpage

\begin{proposition}
The explicit formula for the $\oplus$-sum \eqref{eq:defoplus} of two velocities $\mathbf v, \mathbf u\in \mathbf T^\bot$
is  
\begin{equation}
\label{eq:add5}
\boxed{
\begin{array}{cc} \hbox{\small \sf Velocity addition}\\[-3pt] \hbox{\small \sf formula nr 5} \end{array} \qquad\qquad
\mathbf v  \oplus \mathbf u
= \frac{  \left(1 - \frac{\langle \mathbf v, \mathbf u\rangle}{1+\delta}\right) \, \mathbf v+ \delta \mathbf u  }
{1 -   \langle\mathbf v, \mathbf u\rangle}
\qquad~~
}
\end{equation}
It has the following meaning:
\begin{enumerate}
\item
If an object $B$ moves with respect to $A$ with velocity $\mathbf v$,
this means that the chronor of $B$ is proportional to $\mathbf T+\mathbf v$.  
More precisely, it is its normalized version $\mathbf T' = (\mathbf T+\mathbf v)/|\mathbf T+\mathbf v|$.
Clearly, $\pi_{\mathbf T}(\mathbf T')=\mathbf v$. \\[-19pt]
\item
If object $C$ moves with respect to $B$ with some velocity $\mathbf u'$, this means that the chronor $\mathbf T''$ of $C$ is
proportional to  $\mathbf T'+\mathbf u'$.
Clearly, $\pi_{\mathbf T'}(\mathbf T'') = \pi_{\mathbf T'}(\mathbf T'+\mathbf u')$.  \\[-19pt]
\item 
Object $C$ moves with respect to $A$ with velocity $\mathbf v\oplus u$, which  
is the pseudo-project $\mathbf T''$ onto space $\mathbf T^\bot$.
\end{enumerate}

\end{proposition}

\noindent
{\bf Proof:}  Substitute $\mathbf u'$ to \eqref{eq:add4}.
Use the identity  $1+\|\mathbf v\|^2 = \delta^2$ a couple of times. 
\QED

~\\
To bring it closer to the standard textbook form, replace the inner product by the Euclidean metric, i.e., 
$\langle \mathbf a,\, \mathbf b \rangle = -\mathbf a \cdot \mathbf b$. 
Then the above reads:
\begin{equation}
\label{eq:add55}
\boxed{
\begin{array}{cc} \hbox{\small \sf Velocity addition}\\[-3pt] \hbox{\small \sf formula nr 5'} \end{array} \qquad\qquad
\mathbf v  \oplus \mathbf u
= \frac{\left(1 + \frac{\mathbf v\cdot \mathbf u}{1+\delta}\right) \, \mathbf v      +    \delta \mathbf u    }
{1 +  \mathbf v\cdot \mathbf u}
\qquad~~
}
\end{equation}
For comparison, here is the formula taken from \cite{Mo}:
\begin{equation}
\label{eq:Moller}
\boxed{
\begin{array}{cc} \hbox{\small \sf ~M{\o}ller's }\\[-3pt] \hbox{\small \sf formula}\\ ~~\end{array} \qquad\qquad
\mathbf v\oplus \mathbf u   
= \frac{    \sqrt{1-|\mathbf v|^2} \, \mathbf u  +
                    \left( \frac{ \left( 1-\sqrt{1-|\mathbf v|^2} \right)           \, \mathbf v\cdot \mathbf u }
                                {|\mathbf v|^2}
                                                        +1 \right)\, \mathbf v
}
      {1 + \mathbf v\cdot \mathbf u}
\qquad~~
}
\end{equation}
The reader may check equivalence of the last two formulas
 (the version presented in \eqref{eq:add55} seems somewhat easier on the eyes).

~

\noindent
{\bf Problem:} (Poincar\'e formula) Express \eqref{eq:add5} as 
$$
\mathbf v  \oplus \mathbf u
= \frac{\mathbf v +\mathbf u - \mathbf \varepsilon  }
{1 -   \langle\mathbf v, \mathbf u\rangle}
\quad 
\hbox{where }
\quad
\mathbf \varepsilon =  (1-\delta)\,  \mathbf u  +  \frac{\langle \mathbf v,\mathbf u\rangle}{1+\delta} \, \mathbf v  \,.
$$
Show implication: \   $\mathbf v|| \mathbf u \ \Rightarrow \ \varepsilon = 0$.
\\

The map introduced in Proposition \ref{thm:turn} may be described in a more geometrically sound way as follows:
\newpage
\begin{proposition}
\label{thm:turn2}
Let $\mathbf T_1$ and $\mathbf T_2$ be two chronors (unit time-like future-oriented vectors) in Minkowski space $M$.
Define $\mathbf S = \mathbf T_1 + \mathbf T_2$
and a map
$$
\xi:  M\to M : \ \quad 
\mathbf v \ \mapsto  \ 
\mathbf v - 2\frac{ \langle \mathbf S,\, \mathbf v\rangle }{\langle\mathbf S,\, \mathbf S\rangle } \mathbf S
$$
is an isometry.  Moreover, it is an isometry between the two private spaces
$$
\xi : \mathbf T_{1}^\bot \to \mathbf T_2^\bot ,  \qquad\hbox{and}\qquad \ \xi : \mathbf T_{2}^\bot \to \mathbf T_1^\bot ,
$$
Also, the length of the sum of the chronors is related to the mutual speed $v$ of the observers: 
$$ 
 |\mathbf S |^2  = {\langle\mathbf S,\, \mathbf S\rangle }  = 2 \frac{\delta+1}{\delta}
\qquad \delta = \sqrt{1-v^2}
$$
where $v^2 = |\mathbf v_{12}|^2 = |\mathbf v_{21}|^2$ and $\delta = \delta_{12}$.
\end{proposition}

\section{Adding velocities via Lorentz group}

Let us remark, just for completeness, on  the group-theoretic formulation of velocity addition.
In essence it is the formula \eqref{eq:defoplus} in which the map $\xi$ s extended to the whole Minkowski space $M$.

Let $\mathbf A$ and $\mathbf B$ be two future-oriented time-like vectors.
They define a Lorenz transformation $G_{\mathbf A,\mathbf B}\in \Lambda$ such that
$$
\begin{array}{ll}
1. &G_{\mathbf A,\mathbf B}(\mathbf A)\wedge \mathbf B = 0
        \qquad\hbox{($\mathbf B$ is collinear with the image of $\mathbf A$)}\\[7pt]
2. &G_{\mathbf A,\mathbf B} (\mathbf x) = \mathbf x \ \hbox{for every} \ \mathbf x\in (\mathbf A\wedge \mathbf B)^\bot
\end{array}
$$
Any vector $\mathbf v\in \mathbf T^\bot$, $|\mathbf v|>-1$, defines uniquely a  
transformation 
$$
G_{\mathbf T; \mathbf v} = G_{\mathbf T, \mathbf T+\mathbf v} 
$$
 called a {\bf boost} along velocity $\mathbf v$ in the context of $\mathbf T$.
(Note the semicolon in the subscript.)

Now, define the ``sum of velocities'' as a map
$$
\oplus:\ \mathbf T^\bot\times\mathbf T^\bot \longrightarrow \mathbf T^\bot
$$ 
which for two vectors $\mathbf v, \mathbf w \in \mathbf T^\bot$ is 
\begin{equation}
\label{eq:add6}
\boxed{
\begin{array}{cc} \hbox{\small \sf Velocity addition}\\[-3pt] \hbox{\small \sf formula nr 7} \end{array} \qquad\qquad
\mathbf v\oplus \mathbf w = \pi_{\mathbf T} \left (  
G_{\mathbf T ;\mathbf v} (\mathbf T+ \mathbf u)
                          \right )\
\qquad~~
}
\end{equation}

\noindent
This follows directly from the substitution:
$$
\begin{array}{lll}
\mathbf T_3 &\sim& \ooverline{\mathbf  T +\mathbf v}+ \mathbf w\\
&=& G_{\mathbf T; \mathbf v} (\mathbf T) + G_{\mathbf T; \mathbf v}(\mathbf u)\\
&=& G_{\mathbf T; \mathbf v} (\mathbf T+ \mathbf u)
\end{array}
$$
Now, to read off the relative velocity, apply the pseudo-projection $\pi_{\mathbf T}$.
\QED


~\\
{\bf Remark on the rotation.}
Composition of the two respectful boost, $A\to B$ and $B\to C$,  does not coincide with a single 
simple boost $A\to C$.  Instead, we have the well-known fact that
a composition of simple boosts can be decomposed into a composition of a single boos (along $v\oplus w$
and a simple space-like rotation.
We leave it at that as it is not the main point of these notes.
%


\newpage

\section{Adding velocities via menhir loop}

Here we summarize yet another formula for addition of velocities which results from approach \cite{jk-cromlech}.
Two vectors $\mathbf v$ and $\mathbf w$ in $\mathbf T^\bot$ determine a plane $P\subset \mathbf T^\bot$.
As such, it may be given a complex structure
$$
\mathbf T^\bot \ \cong \ \mathbb C
$$
so that $\mathbf v \cdot \mathbf v = \bar{\mathbf v}\mathbf v$, where the bar denotes complex conjugation.
Let us define a map which is a nonuniform reversible scaling: 
$$
\begin{array}{llll}
\mu:        &\mathbb C \ \to \ \mathbb C
&\quad
v \mapsto \mu(v)\equiv  \tilde z &= \displaystyle\frac{z}{1-\sqrt{1-|z|^2}}
\\[11pt]
\mu^{-1}: &\mathbb C \ \to \ \mathbb C
&\quad
z \mapsto  \mu^{-1}(z) &= \displaystyle\frac{2z}{1+|z|^2}
\end{array}
$$
where $|z|^2 = \bar z z$.
If $\mathbf v\in \mathbb C$ represent a velocity, its image $\mu(\mathbf v)$ is called the {\bf menhir} associated to $\mathbf v$.
As it can be checked, $\mu\circ\mu^{-1} = {\rm id} = \mu^{-1} \circ\mu $.
Define a ``{\bf menhir loop}'' as an algebra $(\mathbb C, \boxplus)$ where:
\begin{equation}
\label{eq:addbox}
a\boxplus b = \frac{a+b}{1+\bar a b}
\end{equation}
(For the case of vanishing denominator one needs to use a limit.)
Here is the claim:  The relativistic composition of velocities 
is an altered menhir product:
\begin{equation}
\label{eq:add77}
\mu(\mathbf v\oplus \mathbf u) = \mu(\mathbf v) \boxplus (\mathbf u)
\end{equation}
Hence a new formula for the addition:
\begin{equation}
\label{eq:add7}
\boxed{
\begin{array}{cc} \hbox{\small \sf Velocity addition}\\[-3pt] \hbox{\small \sf formula nr 7} \end{array} \qquad\qquad
\mathbf v\oplus \mathbf u = \mu^{-1}(\tilde {\mathbf v} \boxplus \tilde {\mathbf u})
\qquad~~
}
\end{equation}

~

For the proof see \cite{jk-cromlech}.  The above equation is simple enough to serve 
the standard way to perform such addition.
%
The relativistic composition of {\it collinear} velocities,  
has a simple algebraic form discovered in 1904 by Henri Poincar\'e \cite{Po}:
\begin{equation}
\label{eq:poincare}
v \voplus w =   \frac{v+w}{1+vw} \,.
\end{equation}
%

As already mentioned, the product $\oplus$ is in general not commutative and not associative,
hence it does not define a group but rather a loop (quasigroup with identity), the {\bf M{\o}ller loop}.
Similar properties of $\boxplus$ define on $\mathbb C$ the ``{\bf menhir loop}''. 
We may view the map $\mu$ as an isomorphism of loops, the M{\o}ller loop and the menhir loop: 
$$
(\mathbb C,\, \oplus) \ \longrightarrow \ (\mathbb C,\, \boxplus )
$$
where the map $\mu$ transforms the rather unpleasantly involved 
standard formula \eqref{eq:Moller} 
into a simple product \eqref{eq:addbox},
the form of which is very similar to the 1-dimensional Poincar\'e formula \ref{eq:poincare},
except the context of the complex algebra and the conjugation in the denominator.

\begin{figure}[h]
     \centering
    \includegraphics[scale=.71]{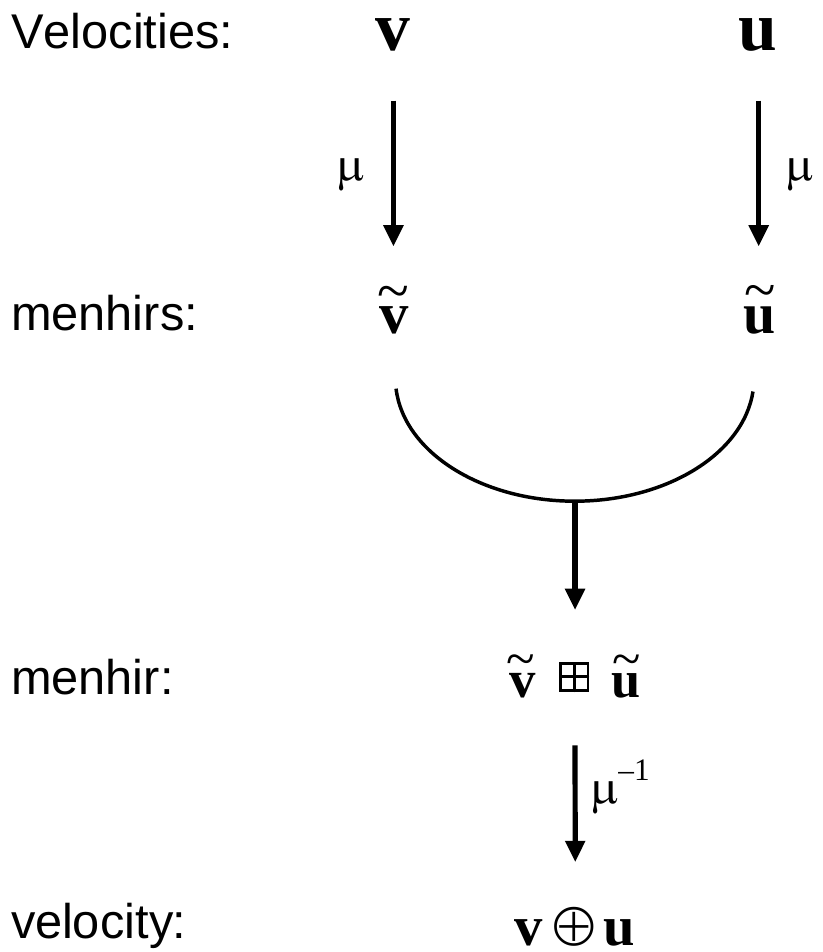}
     \caption{\small Velocity addition as an altered menhir loop.}
     \label{fig:intro2}
\end{figure}

The equivalence of the menhir formulation \eqref{eq:add7} of the velocity addition with the standard \eqref{eq:Moller} 
can be established by direct calculation but it is not entirely effortless.
It seems more reasonable to derive \eqref{eq:add7} from the scratch, 
by geometric analysis of the celestial sphere (projective version of the isotropic cone).
The addition of velocities is defined as follows:
One first defines a observer as an isometric embedding of an $n$-dimensional Euclidean space into the Minkowski space, 
$\lambda: E\to M \cong \mathbf R^{1,n}$.
Space $E$ may be understood as a ``lab'',  and $\lambda(E)$ as its instance in $M$.
The addition of velocities is performed in $E$.  Let $\mathbf v_1, \mathbf v_2\in E$ be two vectors of norm less than 1.
Plan the order of events:  first, accelerate the object along vector $\mathbf v_1$, or rather $\lambda(\mathbf v_1)$.
This  leads to a new embedding of the Euclidean space, say $\lambda': E\to M$.
In the next step, accelerate the lab by the new image of $\mathbf v_2$ 
in the Minkowski space, i.e., by $\lambda'(\mathbf v_2)$.
Thus the claim is that the two accelerations can be replaced by one, namely 
$\mathbf v_1\oplus \mathbf v_2$.  
For details, see \cite {jk-cromlech}.

%
%

\section{Non-associativity of the velocity loop}

\begin{figure}[h]
\large
$$
\begin{tikzpicture}[baseline=-0.7ex]
    \matrix (m) [ matrix of math nodes,
                         row sep=3.5em,
                         column sep=4em,
                         text height=3ex, text depth=2ex] 
 {
  \quad \mathbf T_0\quad &\quad \mathbf T_1 \quad    &   \quad \mathbf T_2 \    & \quad \mathbf T_3  \quad\\
     };
    \path[-stealth, thick]
       (m-1-1) edge node[below] {$\mathbf v_{01}$}  (m-1-2)
       (m-1-2) edge node[below] {$\mathbf v_{12}$}  (m-1-3)
       (m-1-3) edge node[below] {$\mathbf v_{23}$}   (m-1-4)
       (m-1-1) edge [out=-40, in=-140]   node[below] {$\mathbf v_{02} $}       (m-1-3)
       (m-1-2) edge [out=-40, in=-140] node[below] {$\mathbf v_{13}$}    (m-1-4)
       (m-1-1) edge [out=30, in=150]   node[above] {$\mathbf v_{03}$}  (m-1-4)
;
\end{tikzpicture}   
$$
\caption{Notations for the velocities among four objects / observers}
\label{fig:vij}
\end{figure}
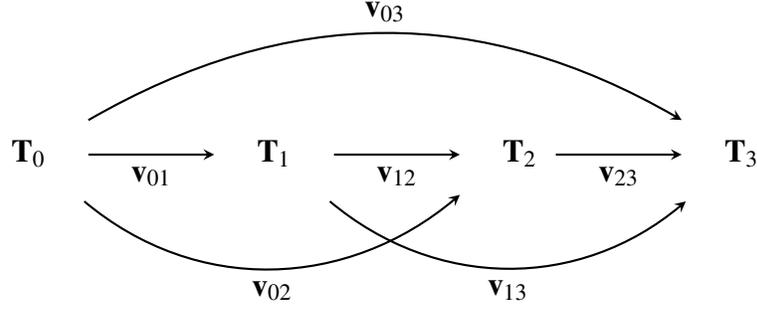

Suppose we have four observers (objects, reference labs) with the corresponding chronors indexed 0 through 3.
The mutual velocities are portrayed in Figure \ref{fig:vij}. 
(Think of $\mathbf T \equiv \mathbf T_0$ as your home, 
$\mathbf T_1$=train, $\mathbf T_2$=turtle in the train, $\mathbf T_3$= snail on the turtle).
Let us denote the mutual velocities as follows: 
$$
\mathbf v_{ij} = \pi_j(\mathbf T_i) = \hbox{(velocity of object {\it j} as measured in the space of {\it i}}) \ \  \in \mathbf T_i^\bot
$$  

\noindent
The task is to recover the velocity of the fourth object (snail) with respect  to the first (home)
from the chain of mutual velocities (see Figure \ref{fig:vij}). 

%
%
%

~

\noindent
The answer is simple
$$
\mathbf v_{03} \ = \      \mathbf v_{01} \; \hhat\oplus \;    \mathbf v_{12}  \;\hhat\oplus\;   \mathbf v_{23} 
$$
There are no brackets on the right side since the ``hat'' addition (Formula 4) 
is associative as the direct descend of the obviously associative Formula 1.  
That is,  
\begin{equation}
\label{eq:hatass}
\boxed{\qquad
 ( \mathbf v_{01} \; \hhat\oplus \;    \mathbf v_{12})  \;\hhat\oplus\;   \mathbf v_{23} 
\quad   =  \quad 
   \mathbf v_{01} \; \hhat\oplus \; ( \mathbf v_{12} \; \hhat \oplus \;  \mathbf v_{23} )
\qquad\phantom{\Big|}}
\end{equation}
The problem starts when we attempt to represent each velocity by  vectors ``at home'', in $\mathbf T_0^\bot$. 
We do it by the map described in Proposition \ref{thm:turn2}  (used before in the form of Eq. \eqref{eq:turn})
denoted:   
%
%
\begin{equation}
\xi_{ij}: \ \mathbf T^{\bot}_j \ \to \ \mathbf T^{\bot}_i 
\end{equation}
This map (whose the exact algebraic form is not essential for now) allows us 
to express the addition as pairing vectors from the same space, i.e.:
$$
 \mathbf v_{ij}\  \hhat\oplus \ \mathbf v_{jk} \ = \ \mathbf v_{ij} \ \oplus \ \xi_{ij} \mathbf v_{jk} 
$$   
(Actually this is the very definition of $\oplus$.)
Let us see what happens with the associativity \eqref{eq:hatass}:
%
{\large
$$
\begin{tikzpicture}[baseline=-0.7ex]
    \matrix (m) [ matrix of math nodes,
                         row sep=2.5em,
                         column sep=.17em,
                         text height=3ex, text depth=2ex] 
 {
  \left(\phantom{\big|\!\!}\right. \mathbf v_{01} &\hhat\oplus &    \mathbf v_{12} \left. \phantom{\big|\!\!}\right)  &\hhat\oplus &  \mathbf v_{23} 
&  \ \ \  = \ \ \  &
   \mathbf v_{01} &\hhat\oplus &   \left(\phantom{\big|\!\!}\right. \mathbf v_{12} &\hhat\oplus &  \mathbf v_{23} \left. \phantom{\big|\!\!}\right) 
     \\
  \left(\phantom{\big|\!\!}\right. \mathbf v_{01} & \oplus &    \xi_{01} \mathbf v_{12}\left. \phantom{\big|\!\!}\right)  & &   
&   &
                           &           &  \left(\phantom{\big|\!\!}\right. \mathbf v_{12} &\oplus & \xi_{12}  \mathbf v_{23} \left. \phantom{\big|\!\!}\right) 
   \\
   \left(\phantom{\big|\!\!}\right. \mathbf v_{01} &\oplus &    \xi_{01} \mathbf v_{12}) &\oplus &   \xi_{02}\mathbf v_{23} 
&  =  &
   \mathbf v_{01} &\oplus &  \xi_{01}  \left(\phantom{\big|\!\!}\right. \mathbf v_{12} &\oplus & \xi_{12}  \mathbf v_{23} \left. \phantom{\big|\!\!}\right) 
   \\
     };
    \path[-stealth, thick]
       (m-1-1) edge node[right] {$$}  (m-2-1)
       (m-1-3) edge node[right] {$\xi_{01}$}  (m-2-3)
       (m-1-9) edge node[right] {$$}  (m-2-9)
       (m-1-11) edge node[right] {$\mathbf \xi_{12}$}  (m-2-11)

       (m-1-5) edge node[below=20pt] {~~~~~~~~$\xi_{12}$}  (m-3-5)
       (m-1-7) edge node[below=20pt] {~~~~~~~~}  (m-3-7)
       (m-2-2) edge node[right] {$$}  (m-3-2)
       (m-2-10) edge node[right] {$\xi_{01}$}  (m-3-10)
       (m-1-9) edge node[right] {$$}  (m-2-9)
       (m-1-11) edge node[right] {$\mathbf \xi_{12}$}  (m-2-11)

;
\end{tikzpicture}   
$$
}
Distributing the $\xi_{01}$ in the last bracket we get the following:
$$
\boxed{
\qquad
 \left(\phantom{\big|\!\!}\right. \mathbf v_{01} \oplus     \xi_{01} \mathbf v_{12} \left. \phantom{\big|\!\!}\right) \oplus    \xi_{02}\mathbf v_{23} 
\quad   =  \quad 
   \mathbf v_{01} \oplus  \left(\phantom{\big|\!\!}\right.\xi_{01}  \mathbf v_{12} \oplus \xi_{01} \xi_{12}  \mathbf v_{23}  \left. \phantom{\big|\!\!}\right)
\qquad\phantom{\Bigg|}
}
$$   
which explains why $\oplus$ (defined on $\mathbf T^\bot$ is {\bf not} associative.
%
%
%
%
%
%
%
If we denote the images of the mutual velocities in the home space $\mathbf T^\bot$ as
$$
\widehat{\mathbf v}_{12} =\xi_{01}\mathbf v_{12}
\qquad 
\widehat{\mathbf v}_{23} =\xi_{02}\mathbf v_{23}
\qquad
\widehat{\widehat{\mathbf v}}_{23} =\xi_{01}\xi_{12}\mathbf v_{12}
$$
then the above form of ``associativity'' becomes
$$
 ( \mathbf v_{01} \;\oplus \;    \widehat{\mathbf v}_{12} ) \;\oplus\;    \widehat{\mathbf v}_{23} 
\quad   =  \quad 
   \mathbf v_{01}\; \oplus \; ( \widehat{\mathbf v}_{12} \;\oplus \; \widehat{\widehat {\mathbf v}}_{23} )
$$   
with all vectors from the {\it same} private space $\mathbf T^\bot$.
\\

\noindent{\bf Analysis:}
The reason for non-associativity lies in the fact that $\widehat{\mathbf v}_{23}\not= \widehat{\widehat{\mathbf v}}_{23}$, or,
in general:
$$
\xi_{02}\ \not= \ \xi_{01}\,\xi_{12}
$$
The equality happens only if chronors $\mathbf T_0$, $\mathbf T_1$,  $\mathbf T_2$,  are collinear. 
To stress, each of these maps (extended to the whole $M$) carries the corresponding {\bf chronors}:
$$
\xi_{01}:\mathbf T_1\to \mathbf T_0, \qquad 
\xi_{12}:\mathbf T_2\to \mathbf T_1, \qquad \xi_{02}:\mathbf T_2\to \mathbf T_0
$$
but not necessarily the {\bf other vectors} considered.  
~\\

\begin{figure}[h]
     \centering
    \includegraphics[scale=.8797]{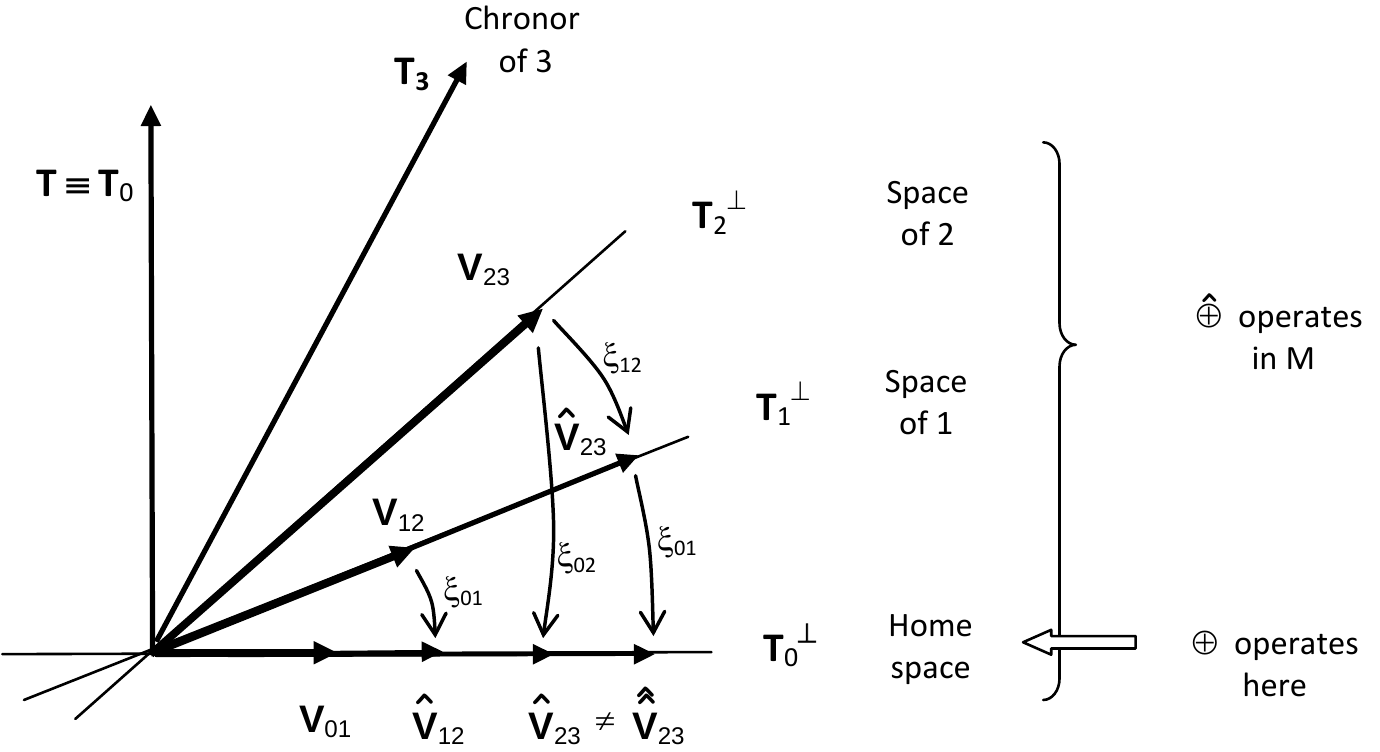}
     \caption{\small Bringing    velocities down to $\mathbf T^\bot$ (not to scale).}
     \label{fig:ksi}
\end{figure}

~\\
\noindent
{\bf Analogy:}  To nail the mechanism that brings about the nonassociativity, 
consider the family of affine maps $\mathbb R \to \mathbb R$ of the form $x\mapsto a\!\cdot\! x+b$.
Any two points $p,q\in\mathbb R$ determine a unique map $f_{qp}$ of the special form $x\mapsto a\!\cdot\! x+1$ that carries  $p$ to $q$.
Let us look at an example:\vspace{-7pt}
$$
\begin{tikzpicture}[baseline=-0.7ex]
    \matrix (m) [ matrix of math nodes,
                         row sep=3.7em,
                         column sep=4.5em,
                         text height=3ex, text depth=2ex] 
 {
  \quad 0\  &\  1 \     &   \ 2 \   & \ 3  \ & \ 4 \  &  ~\\
     };
    \path[-, thick]
       (m-1-1) edge  (m-1-2)
       (m-1-2) edge  (m-1-3)
       (m-1-3) edge  (m-1-4)
       (m-1-4) edge  (m-1-5)
       (m-1-5) edge  (m-1-6);
    \path[->]
       (m-1-2) edge [out=40, in=140]   node[above] {$f_{21} $}       (m-1-3)
       (m-1-3) edge [out=40, in=140] node[above] {$f_{32}$}    (m-1-4)
       (m-1-2) edge [out=60, in=120]   node[above] {$f_{31}$}  (m-1-4)
;
\end{tikzpicture}   
$$
It is easy to determine each ``$a$''
$$
f_{21} (x) = 1\!\cdot\! x+1
\qquad
 f_{32} (x) = 1\!\cdot\! x+1
\qquad
f_{31} (x) = 2x+1
$$
When applied to $x=1$, we have $f_{32}f_{21} (1) = f_{31}(1)$, but in general
$$
f_{32}\circ f_{21} (x) = x+2 \qquad \not\equiv \qquad  f_{31}(x)= 2x+1
$$

\noindent
{\bf A closer analogy:} Consider rotations in the Euclidean space $\mathbb R^3$.  
Take two books in the same position laying flat  in front of you on the table, with an arrow sticking out in the {\bf x}-direction. Then do:
\begin{description}
\item{\bf Book1:} Turn it by $90^\circ$  so that the arrow goes from {\bf x}-axis to {\bf y}-axis, and then by  $90^\circ$
so that the vector goes from {\bf y}-axis to {\bf z}-axis
\item{\bf Book 2:} Turn it book by  $90^\circ$ so that the arrow goes from {\bf x} axis directly to {\bf z} axis.
\end{description} 
Both operations carry the vector to the same position, but the orientations of the books differ.
This analogy is close to the relativistic considerations of velocities.  
The hyperbolic rotations follow the same rules as the regular rotations.

\begin{figure}[h]
     \centering
    \includegraphics[scale=.7]{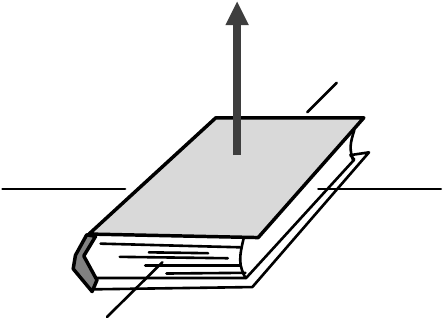}
     \caption{\small A book to play with.}
     \label{fig:book}
\end{figure}

\begin{figure}[h]
     \centering
\hrule

~\\ ~

    \includegraphics[scale=.73]{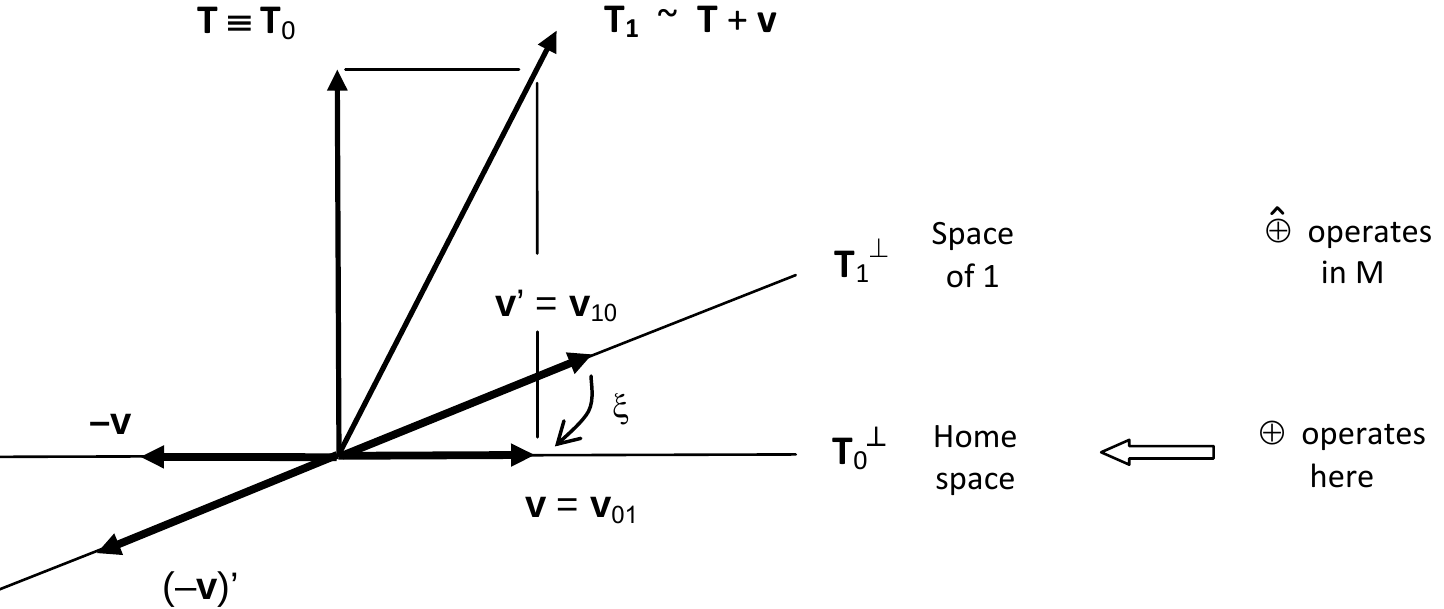}
   
~\\  

\hrule

~

  \caption{\small Reciprocal velocities.}

     \label{fig:minus}
\end{figure}

\noindent
{\bf Reciprocity of velocities.} 
Yet another popular confusion concerns reciprocal velocities.  
If object $B$ moves with respect to object $B$ with velocity $\mathbf v$,
one cannot conclude that object $A$ moves with respect to object $B$ with velocity $(-\mathbf v)$,
These vectors lie actually in in different subspaces of $M$, different `private spaces''.
Here is the situation presented in terms of symbols used in these notes.
See Figure \ref{fig:minus}.

%
%

\begin{equation}
\begin{array}{lll}
\mathbf v \oplus (-\mathbf v) &= 0 \qquad \hbox{and} \qquad &\mathbf v+(-\mathbf v)  \ = 0\\  
\mathbf v \; {\hhat\oplus} \; (-\mathbf v)' &= 0 \qquad \hbox{but} \qquad &\mathbf v+(-\mathbf v)' \not= 0  
\end{array}
\end{equation}
where 
$$
\mathbf v =\mathbf v_{01} \qquad \hbox{and}\qquad (\mathbf v)'=\xi \mathbf v 
$$
Simple calculation gives: 
$$
\mathbf v_{10} \ = \ (-\mathbf v)' \  = \ -\mathbf v'  \ = \ \frac{-\mathbf v +\|\mathbf v\|^2 \mathbf T}{\delta} 
$$

\section{Summary}

The so-called addition of velocities formula is a simple consequence 
of the geometry of the Minkowski space.
It has two types of expressions:

\begin{enumerate}

\item
If velocities are defined as the mutual configurations of two observers, they are space-like vectors in the private space of the first 
observer.  In this case the addition formula performed with $\hhat\oplus$ is associative.  
But the reciprocity has a less direct form:
$$
\begin{array}{lll}
\hbox{associativity:}\quad 
&(\mathbf v_{01}\ \hhat\oplus \ \mathbf v_{12}) \  \hhat\oplus \ \mathbf v_{23} 
\ = \ \mathbf v_{01}\ \hhat\oplus\ (\mathbf v_{12}\ \hhat\oplus \  \mathbf v_{23} ) \\[7pt]
\hbox{reciprocity:}\quad 
&\mathbf v_{01} \ \hhat\oplus \ \mathbf v_{10} \not= 0 
\end{array}
$$
(in general  $\mathbf v_{01} \not=- \mathbf v_{10}$,
and actually $\mathbf v_{01} + \mathbf v_{10} \sim \mathbf T_0+\mathbf T_1 $). 

\item
If all velocities are brought to a single space by appropriate map ($\xi$),
e.g., to the private space $\mathbf T^\bot$ of the first observe,  
then the $\hhat\oplus$-product must be replaced by $\oplus$. 
The simple three-terms form of associativity is gone for such images, although the ```reciprocity'' 
is now obeyed:
$$
\begin{array}{lll}
\hbox{associativity:}\quad 
&(\mathbf v_{01}\ \oplus \ \widehat{\mathbf v}_{12}) \  \oplus \ \widehat{\mathbf v}_{23} 
\ \not= \ \mathbf v_{01}\ \oplus\ (\widehat{\mathbf v}_{12}\ \oplus \  \widehat{\widehat{\mathbf v}}_{23} ) \\[7pt]
\hbox{reciprocity:}\quad 
&\mathbf v_{01}\oplus \widehat{\mathbf v}_{10} = 0 \qquad\hbox{and}\qquad 
\mathbf v_{01} + \widehat{\mathbf v}_{10} = 0 
\end{array}
$$
Because of different paths of bringing the velocity to $\mathbf T^\bot$, the last entry has two images, 
$\widehat{\mathbf v}_{23} \not= \widehat{\mathbf v}_{32}$.
Therefore by just taking three vectors $\mathbf a,\mathbf b,\mathbf c\in \mathbf T^\bot$ 
one should not expect associativity if $\oplus$ is understood ``arithmetically'', 
i.e. without the observers' context.
Thus in general:
$(\mathbf a\oplus \mathbf b)  \oplus \mathbf c 
\ \not= \ \mathbf a \oplus\ (\mathbf b \oplus  \mathbf c) 
$

\end{enumerate}

The menhir calculus refers to the second type of addition and has surprisingly simple form.
Additionally, it has a simple geometric representation.
The connection with the standard addition is 
$$
\lambda (\mathbf v_1\dot\oplus\;\mathbf v_2) \ = \ \lambda (\mathbf v_1) \oplus \lambda (\mathbf v_2)
$$
where $\dot\oplus$ denotes the addition defined via menhir formulation, the regular $\oplus$ denotes 
the standard M{\o}ller's formula,
and $\lambda:E\to M$ is an instance of the lab.

\newpage 

\section*{Appendix: The structure of the special theory of relativity}

The special relativity theory (STR) is an exemplary illustration of the power of science: 
it is relatively simple yet profoundly diverges from the common sense. 
Below, a view on the place of the ``velocity paradox'' in the structure of STR is depicted.
\\

\begin{figure}[H]
     \centering
    \includegraphics[scale=.67]{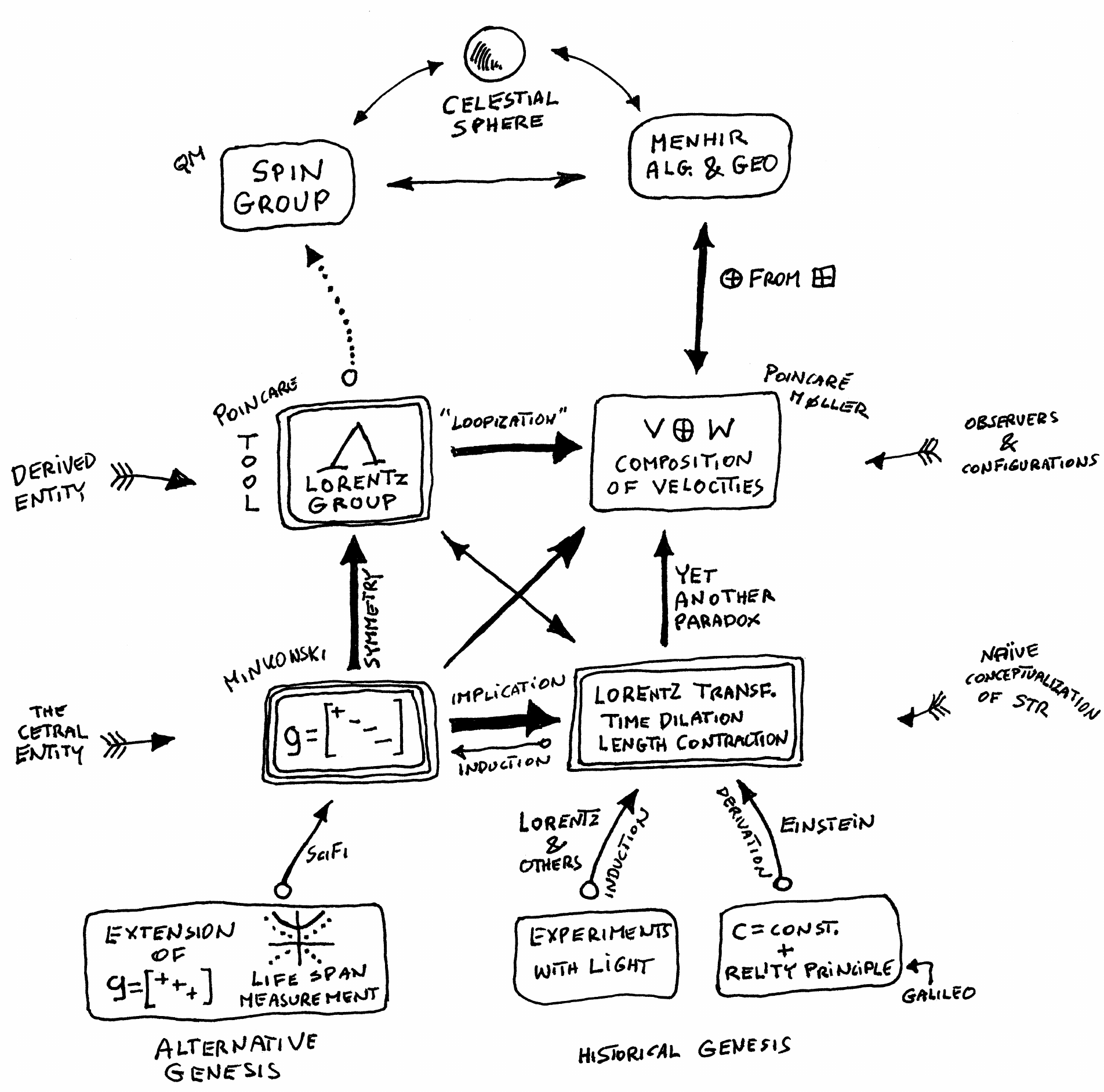}
     \caption{\small Structure of STR.}
     \label{fig:STR}
\end{figure}

The graph in Figure \ref{fig:STR} shows various ingredients of the theory and their mutual relations
(and clearly does not need to be universally shared).
We will walk through it.

Let us start with the {\bf four central boxes} connected by the thick arrows.

\begin{enumerate}

\item
{\bf Three-line box:}
The very central ingredient of the STR is the geometry of  {\bf inner product} of the Minkowski space.
It is located in the above diagram as the triply boxed panel.
This geometry determines the other features.

\item
{\bf Upper double line box:}
The {\bf Lorenz group} is  the symmetry group of the inner product. 
It is an expedient tool for deriving other properties 
and therefore is often introduced as the central concept of STR
while it should be viewed as derived from $g$.
(misleadingly).

\item
{\bf Double line box to the right:}
Basic ``paradoxes'':  Lorentz contraction and time dilation can be simply explained geometrically. 
They are paradoxes only in the sense they do not agree with the ``intuitive'' Galilean world-view.
Historically, the equations of STR were deduced from experiments within Galilean-Newtonian model 
and therefore seemed puzzling. 

\item
{\bf Single line box to the right:}
The paradox of velocity additions is yet another conundrum within the Galilean context. 
It may be derived from the equations of the``Lorentz paradoxes" or from the Lorenz group, as it is usually done nowadays.
But it may also be derived directly from geometry as it was done in the present paper.

\end{enumerate}

\noindent
{\bf Bottom line: the origins}

\begin{enumerate}
\item
{\bf Right side:}
Historical origins of the STR lie in the experiments with light:
change of speed in running in water,  and the famous Michelson-Morley experiment.
They lead to the Lorentz equations through mathematical lucky ingenuity.
The 1905 Einstein's paper \cite{Einstein} derives them 
from only two assumptions and initiates maturation of STR.
Other boxes were deduced consequently (Lorenz group by Poincar\'e and  geometry by Minkowski.
The historical arrows in the diagram do not agree with the conceptual arrows.

\item
{\bf Left side:} 
One may speculate a different origin of the STR (distant civilization?).
Suppose one wants to extend the Euclidean metric to the four-dimensional space-time.
Euclidean metric is a mathematicalization of a simple experiment with a rope,
namely extending it in every direction to create a ``circle'', the set of equidistant points.
That geometric shape is sufficient to deduce orthogonality, inner product and norm.
(Euclidean geometry is the first mathematical model of physical space, after all.)

To produce a ``circle'' of equidistant points is space-time, 
replace Euclid's rope by a big number of alarm-clocks,
all set for the same unit of time,
and send them out in different space-time  directions (different speeds).   
Mark the events of alarm clocks going off.
The shape of such equidistant points will reveal the upper piece of a hyperboloid!
This, fortified with some mathematical arguments,  should reproduce the Minkowski metric.  


\end{enumerate}

\noindent
{\bf Upper part of the diagram}

\begin{enumerate}
\item
{\bf Left side:} 
Group theory provides a different representation of the Lie algebra of $\Lambda$,
namely the spin group ${\rm SL}(2,\mathbb C)$.
This remarkable and lucky bonus from mathematics 
turns out to be the essential tool for description of the quantum property of spin!

\item
{\bf Right side:}
An alternative representation of the addition of the velocities is located here.
It may be derived from the conformal maps of the celestial sphere due to boosts.
In its heart it conceals the spin representation.

\end{enumerate}


\end{document}